\documentclass[10pt,conference]{IEEEtran}
\usepackage{braket}

\usepackage{cite}
\usepackage{graphicx}
\usepackage{amsmath,amssymb}
\usepackage{algorithmic}
\usepackage{url}

\usepackage{libertinus}

\usepackage{hyperref}
\begin{document}


\title{Distributed Realization of Color Codes for Quantum Error Correction}
%

\author{
Nitish Kumar Chandra\textsuperscript{1}, 
David Tipper\textsuperscript{1}, 
Reza Nejabati\textsuperscript{2}, 
Eneet Kaur\textsuperscript{2}, 
Kaushik P. Seshadreesan\textsuperscript{1}\textsuperscript{\dag} \\[1ex]
\textsuperscript{1}Department of Informatics \& Networked Systems, School of Computing \& Information,\\
University of Pittsburgh, Pittsburgh, PA 15260, USA\\
\textsuperscript{2}Cisco Quantum Lab, Los Angeles, CA 90404, USA\\[1ex]
\textsuperscript{\dag}Corresponding author: kausesh@pitt.edu
}

\maketitle

\begin{abstract}

Color codes are a leading class of topological quantum error-correcting codes with modest error thresholds and structural compatibility with two-dimensional architectures, which make them well-suited for fault-tolerant quantum computing (FTQC). Here, we propose and analyze a distributed architecture for realizing the (6.6.6) color code. The architecture involves interconnecting patches of the color code housed in different quantum processing units (QPUs) via entangled pairs. To account for noisy interconnects, we model the qubits in the color code as being subject to a bit-flip noise channel, where the qubits on the boundary (seam) between patches experience elevated noise compared to those in the bulk. We investigate the error threshold of the distributed color code under such asymmetric noise conditions by employing two decoders: a tensor-network-based decoder and a recently introduced concatenated Minimum Weight Perfect Matching (MWPM) algorithm. Our simulations demonstrate that elevated noise on seam qubits leads to a slight reduction in threshold for the tensor-network decoder, whereas the concatenated MWPM decoder shows no significant change in the error threshold, underscoring its effectiveness under asymmetric noise conditions. Our findings thus highlight the robustness of color codes in distributed architectures and provide valuable insights into the practical realization of FTQC involving noisy interconnects between QPUs.
\end{abstract}

\begin{IEEEkeywords}
Distributed quantum computing, Color code, Tensor network-based decoder, Concatenated minimum weight perfect matching-based decoder
\end{IEEEkeywords}
\section{Introduction}

Quantum error correction (QEC) is essential for  reliable quantum computation, given the inherent susceptibility of quantum systems to noise caused by decoherence, gate imperfections, and environmental interactions~\cite{nielsen2002quantum, preskill1998reliable, RevModPhys.87.307}. Unlike classical bits, quantum bits (qubits) cannot be copied or measured directly without disrupting the encoded information. Therefore, fault-tolerant methods are required to preserve quantum coherence while detecting and correcting both bit-flip and phase-flip errors~\cite{PhysRevA.52.R2493, gottesman1997stabilizercodesquantumerror}. Although quantum computing offers  speedups for tasks such as factoring, quantum simulation, and optimization~\cite{doi:10.1137/S0097539795293172,10.1145/237814.237866, Montanaro2016}, realizing these advantages at scale depends if logical qubits are robustly protected against physical noise. This is accomplished by encoding logical information across multiple physical qubits and utilizing QEC codes with efficient decoding algorithms and high error thresholds~\cite{lidar2013quantum}.

Among the various quantum error-correcting strategies, topological quantum error-correcting codes—notably surface codes, and color codes have emerged as leading candidates due to their compatibility with physical architectures and scalability. These codes encode logical qubits into large ensembles of physical qubits, allowing quantum information to be protected against Pauli noise through local stabilizer measurements~\cite{KITAEV20032, PhysRevLett.97.180501}. Their inherently local structure simplifies implementation on two-dimensional lattices, while their achievable  fault-tolerance thresholds make them well-suited for near-term and future quantum processors~\cite{google2023suppressing,PhysRevLett.129.030501}. 

  Within the topological code family, color codes are particularly notable for their structural and operational advantages. Defined on trivalent lattices with three-colorable faces, they support transversal implementation of all Clifford gates, thereby simplifying logical operations and enabling low-overhead fault-tolerant protocols~\cite{landahl2011faulttolerantquantumcomputingcolor,PRXQuantum.5.030352}. The 6.6.6 color code, defined on a hexagonal lattice, offers a balanced trade-off between error-correction performance and resource overhead. Recent advances in decoding techniques and logical gate implementations have further improved its suitability for scalable two-dimensional quantum architectures, where low circuit depth and limited interaction range are essential for near-term implementations~\cite{PRXQuantum.5.030352, PhysRevResearch.6.033086}.

The performance and practicality of quantum error-correcting codes (QECCs) depend not only on code distance, locality, and stabilizer layout but also on the efficiency and accuracy of the decoding algorithms used to interpret error syndromes and apply the appropriate corrections~\cite{deMarti_iOlius_2024,campbell2024series}. In practical implementations, decoding is particularly challenging due to the need to correct multiple types of errors, often under stringent time and hardware constraints. Consequently, the development of fast, scalable, and fault-tolerant decoding strategies is essential for the practical realization of quantum computation. To tackle these challenges, various decoding strategies have been proposed. A widely used approach involves projecting the color code onto multiple surface codes, each independently decoded using the Minimum Weight Perfect Matching (MWPM) algorithm, with the results then combined to infer the overall correction~\cite{PhysRevA.89.012317}. This approach capitalizes on the established decoding framework of surface codes while retaining the advantages of color codes. Other methods include Union-Find decoders~\cite{delfosse2021almost}, and tensor network-based techniques decoding~\cite{chubb2021generaltensornetworkdecoding,PRXQuantum.5.040303}. Recently, a matching-based decoder for color codes known as the concatenated MWPM decoder was introduced~\cite{lee2025color}, demonstrating excellent sub-threshold performance. 
Efficient decoding remains a key research challenge, particularly for color codes, which enable a wider set of transversal gates and reduced qubit overhead than surface codes, although surface codes  offer slightly higher error-correcting thresholds~\cite{PRXQuantum.5.030352}.

Achieving large-scale fault-tolerant quantum computation requires architectures capable of implementing quantum error correction across many physical qubits. A promising strategy is the use of modular quantum architectures, in which a large quantum processor is composed of smaller, locally controlled modules such as ion traps, superconducting qubit arrays, or photonic processors—interconnected via quantum communication links~\cite{PhysRevX.4.041041, PhysRevA.89.022317, chandra2024multiplexed}. This modular approach addresses key challenges associated with monolithic designs, including limited qubit connectivity and increased crosstalk between qubits.
Distributing quantum computation across spatially separated modules poses new challenges, particularly in ensuring reliable inter-module communication and managing the increased latency in error detection and correction~\cite{vanmontfort2024faulttolerantstructuresmeasurementbasedquantum,10.1145/3620665.3640388,PhysRevResearch.5.043302}. These links, often implemented via photonic entanglement or matter-light interfaces, can be slow to establish and error-prone. Thus requiring the need for efficient quantum distillation protocols, quantum teleportation, and lattice surgery in-order to perform logical operations across modules~\cite{PhysRevResearch.5.033171,PhysRevResearch.6.043125}. One of the concerns with the inter-module links is that boundary qubits that interface between modules are more susceptible to errors. To ensure fault tolerance in such systems, we must account for asymmetric noise profiles, where qubits associated with interconnects can experience significantly higher error rates than those within a module in the quantum error correction schemes. Our analysis incorporates this asymmetry, which is a critical step in advancing fault-tolerant distributed quantum computation (FTDQC), where noisy quantum channels pose a significant challenge~\cite{PhysRevResearch.5.043302}.



Some recent works have explored the implementation of quantum error-correcting codes for distributed architectures~\cite{Ramette2024,10.1116/5.0200190,sutcliffe2025distributedquantumerrorcorrection}. In Ref.\cite{Ramette2024}, the authors demonstrated that distinct surface code patches can be connected in a fault-tolerant manner, even when the interface is subject to elevated levels of noise. Ref.\cite{10.1116/5.0200190} investigated a distributed toric surface code architecture, where each data qubit resides on a separate node. This study examined how memory decoherence affects the performance of the code and assessed Greenberger–Horne–Zeilinger (GHZ) state generation strategies adapted to varying degrees of decoherence. Ref.~\cite{sutcliffe2025distributedquantumerrorcorrection} proposed hyperbolic Floquet codes as a promising approach to distributed quantum error correction. This work highlighted how the hyperbolic geometry enables efficient encoding of a large number of logical qubits and demonstrated that these codes exhibit strong performance under both local and non-local phenomenological noise.

In this work, we evaluate the error threshold of a $6.6.6$ triangular color code in a modular setting, where the code in one module is scaled to a higher-dimensional code by connecting it to other modules via entangled qubit pairs. Our analysis is performed under a bit-flip noise model that assumes elevated error rates at boundary (seam) qubits, capturing the asymmetry introduced by imperfect quantum interconnects in modular architectures. To assess decoder performance under such asymmetric noise, we perform numerical simulations using two decoding strategies: a tensor network-based decoder and a recently proposed concatenated Minimum Weight Perfect Matching (MWPM) decoder. These decoders were chosen for their complementary strengths—the tensor network decoder offers higher error thresholds, while the concatenated MWPM decoder provides faster computational performance. Our results demonstrate that the color code maintains fault tolerance even in the presence of substantial seam noise, highlighting its potential for distributed quantum computing.

The remainder of this paper is organized as follows: Section~\ref{background} provides essential background theory, covering stabilizer quantum error correction, color codes, and the noise model. In Section~\ref{dcc}, we introduce our proposed distributed color code, detailing its theoretical formulation and practical implementation. Section~\ref{plots} presents the results of our simulations, including analyses of logical failure rates and error thresholds under the considered noise model. Finally, Section~\ref{conclusion} also discusses the implications of our findings, their relevance to practical quantum hardware, and potential directions for future work.

\section{Background Theory}\label{background}

\subsection{Quantum Error Correction and Stabilizer Codes}


Quantum error correction (QEC) plays a crucial role in safeguarding quantum information from decoherence and imperfections arising in quantum hardware~\cite{calderbank1996good,Nielsen_Chuang_2010}. A QEC code achieves this by embedding a logical qubit within a subspace of a larger Hilbert space composed of several physical qubits. This subspace is carefully constructed to allow the identification and correction of errors without collapsing or directly measuring the logical quantum state.


The stabilizer formalism, originally developed by Gottesman~\cite{gottesman1997stabilizercodesquantumerror}, offers a robust and elegant framework for formulating and analyzing quantum error-correcting codes. In this approach, a stabilizer code is specified by an abelian subgroup $\mathcal{S}$ of the $n$-qubit Pauli group $\mathcal{P}_n$, with the exception of the element $-I^{\otimes n}$. The associated codespace $\mathcal{C}$ consists of all quantum states that are stabilized by every element of $\mathcal{S}$, meaning each state is a simultaneous $+1$ eigenstate of all stabilizer operators $S_i \in \mathcal{S}$:  \begin{equation}
   \mathcal{C} = \{ |\psi\rangle \in \mathbb{C}^{2^n} \;|\; S_i|\psi\rangle = |\psi\rangle, \forall S_i \in \mathcal{S} \}.
\end{equation}

An important subclass of stabilizer codes is the Calderbank-Shor-Steane (CSS) code~\cite{calderbank1996good, Steane1996}, which is constructed from two classical linear codes \( C_X \) and \( C_Z \), each of length \( n \). Let \( C_X \) be an \([n, k_X, d_X]\) code and \( C_Z \) an \([n, k_Z, d_Z]\) code, such that \( C_X^\perp \subseteq C_Z \) and \( C_Z^\perp \subseteq C_X \). The CSS stabilizer code is then defined by the abelian group generated by the sets:
\[
\mathcal{S}_X = \{ X^{\mathbf{v}} \mid \mathbf{v} \in C_X^\perp \}, \quad
\mathcal{S}_Z = \{ Z^{\mathbf{u}} \mid \mathbf{u} \in C_Z^\perp \}.
\]

This construction enables the independent correction of bit-flip and phase-flip errors and serves as a foundational framework for topological codes such as surface and color codes.


\subsection{Quantum Color Code}

Quantum color codes, proposed by Bombín and Martín-Delgado~\cite{PhysRevLett.97.180501}, constitute an important class of topological stabilizer codes that are particularly well-suited for fault-tolerant quantum computation. They protect quantum information through topological encoding, where logical degrees of freedom are delocalized over the lattice. A distinguishing feature of color codes is their ability to support transversal implementation of a wide range of logical gates, thereby enabling fault-tolerant operations with reduced overhead. These codes are defined on \textit{trivalent lattices} embedded in two-dimensional manifolds, where the faces are \textit{three-colorable}, and physical qubits are located at the vertices.

Each face \( f \) of the lattice is associated with two stabilizer generators, one for each of the Pauli \( X \) and \( Z \) operators, acting on the qubits at the boundary of the face:
\begin{equation}
    S_f^X = \bigotimes_{v \in \partial f} X_v, \quad
    S_f^Z = \bigotimes_{v \in \partial f} Z_v,
    \label{syndrome formula}
\end{equation}
where \( \partial f \) denotes the set of vertices adjacent to face \( f \). The trivalent structure ensures that each qubit is shared by faces of all three colors, allowing it to participate in both X and Z type stabilizers. This structure guarantees the commutativity of the stabilizer group and supports efficient syndrome extraction.

\begin{figure}[hbt!]
    \centering
    \includegraphics[width=0.85\columnwidth]{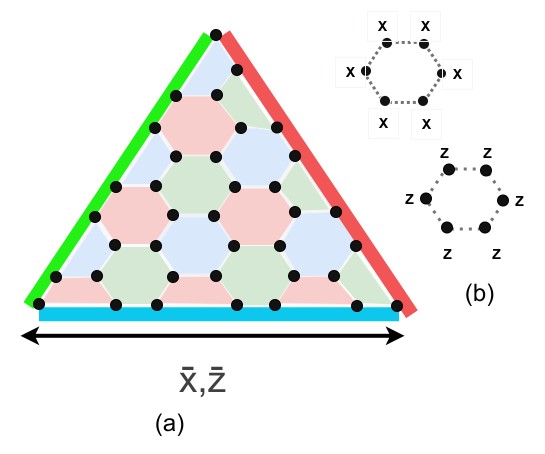}
    \caption{(a) A triangular color code constructed on a hexagonal lattice with three distinct colored boundaries. Physical qubits are positioned at vertices, and each face supports both a \( X \)-type and a \( Z \)-type stabilizer. Logical \( \overline{X} \) and \( \overline{Z} \) operators can be implemented along any of the colored boundaries. The layout shown corresponds to a distance-7 code encoding a single logical qubit. (b) Each face is associated with two stabilizer generators acting on the qubits located at its vertices.}
    \label{fig:plot1}
\label{color_code}
\end{figure}

A well-known instance of color codes is the \((6,6,6)\) code, commonly referred to as the \textit{hexagonal} or \textit{honeycomb color code} (See Fig.~\ref{color_code}). This code is defined on a regular tiling where every face is a hexagon and each vertex has degree three. In the \textit{triangular layout}, the code is restricted to a finite region bounded by three edges, each assigned a distinct color. This configuration encodes a single logical qubit, and logical operators are supported along the corresponding colored boundaries.

\subsection{Noise Model} 

To evaluate the performance of a quantum error-correcting code, it is critical to specify a noise model that reflects the physical error mechanisms. In this work, we consider an independent bit-flip noise model, where each physical qubit is affected by a Pauli-$X$ error independently with probability $p$. The corresponding quantum channel acting on a single-qubit state $\rho$ is given by: 
\begin{equation}
    \mathcal{N}(\rho) = (1 - p)\rho + p\, X \rho X.
\end{equation}

In modular quantum architectures, where quantum processing units (QPUs) are interconnected via photonic or ion-based links, qubits located at module interfaces—referred to as \textit{seam qubits}—are typically more susceptible to noise than bulk qubits. This elevated error rate arises from several practical factors. In particular, nonlocal stabilizer measurements and entangling gates between modules rely on the generation and consumption of Bell pairs over photonic channels. Imperfections in this process, such as photon loss, mode mismatch, and detector inefficiency, introduce additional noise in Bell pairs that can propagate to data qubits during syndrome extraction~\cite{PhysRevA.95.012327}. Moreover, synchronization errors and timing jitter between modules can further degrade the fidelity of inter-module operations~\cite{Burenkov:23}.

To capture this asymmetry, we consider a bit-flip model in which seam qubits experience a higher error probability $p_{seam} > p$, while bulk qubits are subject to the baseline rate $p$. The asymmetric noise model considered here serves as a simplified yet illustrative scenario for analyzing the error-correction threshold of distributed color codes. A more comprehensive and realistic noise model would account for a broader range of error sources, including imperfect state preparation, faulty gate operations, measurement errors, and decoherence at each time step~\cite{PhysRevA.89.022317,PhysRevResearch.7.013313}. Extending the analysis to incorporate these effects will be pursued in future work.

\subsection{Tensor Network-Based Decoding }

Decoding is a critical component of a quantum error correction (QEC) protocol,  as it determines how to interpret the outcomes of stabilizer measurements, commonly referred to as syndromes. This information is then used to apply a suitable recovery operation that restores the quantum state to the protected codespace. Decoding quantum error-correcting codes is particularly challenging due to the degeneracy of error classes, where different error patterns can produce the same syndrome. This indistinguishability hinders accurate error identification and makes exact decoding computationally infeasible for large system sizes, motivating the development of approximate yet scalable decoding strategies~\cite{deMarti_iOlius_2024}.


One promising class of decoding algorithms is based on tensor network methods, which originated from the simulation of quantum many-body systems and have been adapted for decoding tasks, especially for two-dimensional topological codes~\cite{PhysRevX.9.041031}. Prior studies have revealed formal connections between maximum likelihood decoding (MLD) and the statistical mechanics of disordered spin systems, such as random-bond Ising models~\cite{PhysRevA.90.032326}. Within this framework, the decoding task is recast as a problem of probabilistic inference. The objective is to determine the most probable coset of the stabilizer group that contains the actual error, given the observed syndrome. This inference is facilitated by expressing the joint probability distribution of errors and syndromes as a tensor network, where the contraction of the network encodes marginal and conditional probabilities over error configurations~\cite{PhysRevLett.113.030501}.

A particularly effective realization of this approach employs the \textit{Matrix Product State} (MPS) representation~\cite{PhysRevX.9.041031,chubb2021generaltensornetworkdecoding}. In this scheme, the tensor network is contracted sequentially along a linear path through the lattice, forming an MPS. Each tensor encodes information about local error probabilities and stabilizer constraints. The quality of the approximation is governed by the bond dimension \( \chi \), which controls the maximum correlations that the MPS can capture. Higher values of \( \chi \) improve the accuracy of the decoding at the expense of increased computational resources.

MPS-based decoding is especially well-suited to topological codes defined on 2D lattices with local stabilizer structure~\cite{Farrelly_2022}. Its effectiveness stems from its ability to exploit geometric locality and the sparse connectivity of stabilizers. Under the bit-flip noise model, these decoders demonstrate threshold performance up to 10.91\%, close to the theoretical maximum of 10.92\%~\cite{chubb2021generaltensornetworkdecoding}.
Additionally, this method can accommodate a range of noise models, including independent, biased, and spatially correlated noise, by modifying the local tensor components. This adaptability makes it a powerful tool for assessing the performance of quantum error-correcting codes and for achieving high thresholds under practical noise conditions.




\subsection{Concatenated MWPM Decoder}

Minimum-weight perfect matching (MWPM) is one of the most widely used classical decoding algorithms for topological quantum codes, offering both high performance and computational efficiency~\cite{10.1145/3505637,higgott2025sparse}. It has been particularly effective for surface codes, where decoding can be formulated as a graph matching problem. In this approach, syndrome outcomes are interpreted as defect pairs, and the decoder searches for a minimum-weight correction by connecting these defects through the shortest paths on the lattice. Algorithms such as Blossom are commonly employed to compute an optimal matching that minimizes the total correction cost.

However, directly applying MWPM to color codes introduces significant challenges due to their more complex structure. In color codes, each physical qubit is involved in multiple stabilizers. As a result, a single-qubit error can simultaneously violate several checks, leading to syndrome patterns with nontrivial correlations. This complexity makes simple pairwise matching insufficient, as it does not fully capture the dependencies among stabilizer violations. In particular, projection decoders and related variants based on standard MWPM approaches for color codes often exhibit suboptimal logical failure rate scaling of \(p^{d/3}\) below threshold, where \(p\) is the physical error rate and \(d\) is the code distance, compared to the optimal scaling of \(p^{d/2}\)~\cite{PRXQuantum.2.020341,PRXQuantum.3.010310,lee2025color}.
 This unfavorable scaling requires the use of significantly larger code distances to maintain low logical error rates, impacting the overall resource efficiency in fault-tolerant quantum computation.

To address these limitations, the \textit{concatenated MWPM decoder} was recently proposed to  effectively leverage the interplay between geometric structure and stabilizers of color codes~\cite{lee2025color}. The key idea is to perform MWPM in two stages: first on a restricted subgraph of the lattice that excludes one color of stabilizers, and then on an auxiliary graph constructed from the predicted error edges and remaining syndrome data. For example, MWPM is applied to a red-restricted subgraph using violated green and blue stabilizers, producing a set of red edges likely to contain errors. These predicted edges, along with the violated red stabilizers, are used to define a second decoding problem confined to red checks. A second MWPM is then run on this red-only graph to determine the final set of qubits likely to be erroneous. This procedure is repeated independently for each of the three colors, and the correction with the lowest total weight is chosen as the final recovery operation.

One of the main advantages of this decoder is its significantly improved sub-threshold performance, achieving logical error rates that scale as \(\sim p^{d/2}\), which is much closer to the optimal scaling expected for topological codes. This decoder reaches a threshold of 8.2\% when subjected to bit-flip noise.
 As a result, the decoder provides an effective balance between accuracy and computational cost, making it suitable for near-term implementations of fault-tolerant quantum computing architectures.




\section{Distributed Color Code}\label{dcc}

As quantum computing systems grow in size, monolithic architectures encounter fundamental bottlenecks, including limited qubit-to-qubit connectivity, increased  crosstalks, and geometric layout constraints that hinder scalability. These challenges have led to the advancement of \textit{distributed quantum computing}, where multiple quantum processing units (QPUs) are interconnected to perform large-scale computations~\cite{BARRAL2025100747}. This modular approach supports high-fidelity operations within individual QPUs while leveraging photonic links to mediate entanglement and enable gate operations across spatially separated modules~\cite{10835728}.
As a result, distributed quantum architectures offer a promising pathway toward building scalable  fault-tolerant quantum computers.

\begin{figure}[hbt!]
    \centering
    \includegraphics[width=0.95\columnwidth]{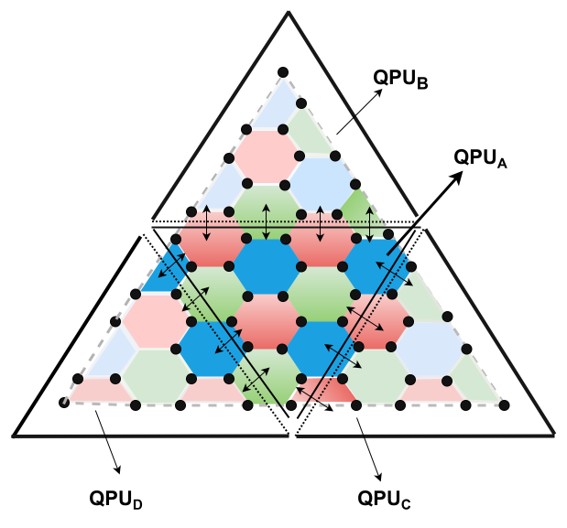}
    \caption{A distance-$9$ color code is distributed across four quantum processing units (QPUs), with a distance-$5$ color code embedded in $QPU_{A}$ and the remaining qubits allocated across the other QPUs. $QPU_{A}$ is connected to the other units via Bell pairs, represented by double arrows (serves as a representative figure, it does not reflect the exact number of Bell pairs). Since the QPUs are spatially separated, nonlocal CNOT gates required for stabilizer measurements can be implemented using these Bell pairs.}
    \label{fig:plot2}

\end{figure}

In this work, we explore a distributed implementation of the triangular 6.6.6 color code, where a logical qubit initially encoded in a lower-distance code within a single QPU is extended to a higher-distance code by incorporating auxiliary qubits distributed across additional QPUs. Equivalently, this can be viewed as partitioning a larger distance color code such that one QPU hosts a smaller distance triangular code, while the remaining physical qubits are allocated to other modules. We illustrate in Fig.~\ref{fig:plot2}, a specific example in which a distance \( d = 9 \) triangular color code is distributed over four QPUs. A distance \( d = 5 \) triangular code is embedded in one QPU, denoted \( QPU_A \), and the additional qubits required to have a higher distance \( d = 9 \)  code are placed in three other QPUs. 

We consider a small triangular color code embedded within \( QPU_A \), which serves as a subregion of a larger-distance code distributed across multiple QPUs. The smaller code is connected to auxiliary qubits in neighboring QPUs via nonlocal inter-module links. As a result, the boundary (or seam) qubits—those along the edges of the triangular code that interface with other modules—are subject to elevated noise due to imperfections in entanglement generation and transmission. We vary the distance of the smaller code, treating it as part of an increasingly larger-distance code, and evaluate the performance of the decoders both with and without elevated error rates applied to the seam qubits.



In color codes, stabilizers are associated with each face, and each stabilizer acts on the qubits surrounding it. These are typically implemented using sequences of CNOT gates between an ancilla and adjacent data qubits. However, in a distributed layout, many of these data-ancilla interactions become nonlocal—requiring CNOTs across QPU boundaries. To implement such inter-module gates, we employ \textit{entanglement-assisted gate teleportation}, a protocol that leverages shared Bell pairs between QPUs to apply nonlocal CNOTs using only local operations, single-qubit measurements, and classical communication~\cite{BARRAL2025100747}.

\begin{figure}[hbt!]
    \centering
    \includegraphics[width=0.99\columnwidth]{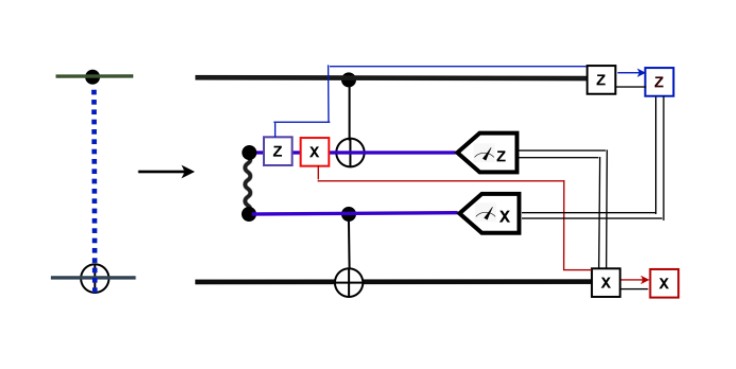}
    \caption{The spatial separation between QPUs necessitates the use of shared Bell pairs to implement nonlocal CNOT gates for stabilizer measurements. Pauli \(X\) and \(Z\) errors affecting a Bell pair (indicated by a black squiggly line) used in a teleported gate (depicted as a blue dotted line on the left side) are propagated to the corresponding qubits involved in the gate operation. Specifically, \(Z\) (phase) errors propagate to the control qubit, while \(X\) (bit-flip) errors propagate to the target qubit of the teleported CNOT gate.}
    \label{fig:plot3}

\end{figure}

A nonlocal CNOT gate can be implemented using a shared Bell pair between the participating QPUs (See Fig.~\ref{fig:plot3}). The protocol begins by distributing a high-fidelity Bell state between communication qubits in the two modules. Local CNOT gates and measurements are then performed to effectively teleport the action of a CNOT from the control qubit in one QPU to the target qubit in the other. Depending on the measurement outcomes, Pauli corrections are applied to complete the operation.

Inter-QPU links, implemented using Bell pairs, are consumed during each round of stabilizer measurements. Seam or boundary qubits, located at the interfaces between QPUs, play a particularly important role in this process. As these qubits are involved in multiple inter-module operations, they are more susceptible to noise introduced by imperfections in the interconnects. To evaluate the impact of this increased noise, we perform simulations using a tensor network-based decoder and a concatenated MWPM decoder, assuming perfect stabilizer measurements. We determine the error correction threshold to assess decoder performance in the presence of noisy inter-module links.

\subsection{Syndrome Extraction} 

In a color code, each face of the hexagonal lattice is associated with two stabilizer measurements: one for the Pauli-\( Z \) operator and one for the Pauli-\( X \) operator (See~Eq.(\ref{syndrome formula}). Each of these \( Z \)-type or \( X \)-type checks is performed using an ancillary qubit placed at the center of the face, which interacts with the six surrounding data qubits through a series of CNOT gates.
\begin{figure}[hbt!]
    \centering
    \includegraphics[width=0.90\columnwidth]{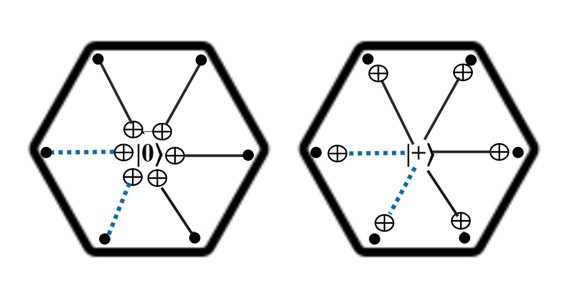}
    \caption{Syndrome extraction circuits for $Z$-type (left) and $X$-type (right) stabilizers in the color code. Each circuit uses an ancilla qubit to interact with six surrounding data qubits via CNOT gates. Two of the CNOTs are nonlocal, indicated by blue dotted lines, while the remaining are local gates confined within a single QPU.}
    \label{fig:plot4}

\end{figure}
As illustrated in Fig.~\ref{fig:plot4}, the circuit on the left shows the process for measuring a $Z$-type stabilizer. The central ancilla qubit is prepared in the $\ket{0}$ state and acts as the target in CNOT gates with control on each of the surrounding data qubits. Once all CNOTs are applied (in a carefully chosen order), the ancilla is measured in the $Z$ basis to obtain the parity of the $Z$ operators on the face. The circuit on the right shows the measurement of an $X$-type stabilizer. This time, the ancilla is initialized in the $\ket{+}$ state and serves as the control of CNOT gates with target on each of the surrounding data qubit. After all gates are executed, the ancilla is measured in the $X$ basis to measure the ($X$-type) checks.

To ensure these gates do not conflict in time or interfere with neighboring operations, each CNOT is scheduled using time labels (not shown), which define the order of execution. This scheduling is important in hardware implementations to avoid simultaneous gate operations on shared qubits. For faces at the boundaries of the code, a similar strategy is followed. While these may involve fewer than six data qubits, the same ancilla-based circuit structure is used, with adjusted timing to reflect the reduced number of interactions~\cite{lee2025color}.

\subsection{Entanglement Overhead across QPU Boundaries}

We consider an architecture in which QPUs are connected via classical communication channels in order to enable the transmission of syndrome measurement outcomes for classical processing. To perform syndrome extraction across two distinct QPUs, we determine the number of Bell pairs required to implement non-local gates between an ancilla qubit on one QPU and data qubits located on the other.

For each stabilizer measurement associated with a hexagonal face at the boundary (e.g., the more prominently shaded blue and red hexagons spanning the interface between $QPU_{A}$ and $QPU_{C}$ in Fig.~\ref{fig:plot2}), the ancilla qubit is placed inside the QPU that hosts the majority of the involved data qubits. For instance, if four out of the six data qubits involved in the stabilizer measurement are located on one QPU, the ancilla is placed within that QPU. This placement minimizes the number of inter-QPU Bell pairs required, reducing it to just two per stabilizer measurement. In cases where the stabilizer acts on an equal number of data qubits across the two QPUs, such as a weight-four stabilizer with two qubits on each QPU, the ancilla qubit can be placed on either QPU. This also requires two Bell pairs to connect the ancilla to the non-local data qubits.

Thus, for a single round of syndrome extraction along the boundary (either $X$- or $Z$-type) in a triangular color code with $k$ data qubits along each boundary edge (see $QPU_{A}$ in Fig.~\ref{fig:plot2}), we require $2(k - 1)$ Bell pairs per edge to connect to the adjacent QPU. Since the triangular layout has three edges, the total number of Bell pairs needed per round of either type is $6(k - 1)$.

\subsection{Noise Propagation}

Imperfect Bell pairs used in teleportation-based CNOT gates can introduce errors that propagate into data qubits during stabilizer measurements. As illustrated in Fig.~\ref{fig:plot3}, Pauli errors on the Bell pair lead to propagation in the teleportation circuit: phase-flip ($Z$) errors affect the control qubit, while bit-flip ($X$) errors propagate to the target qubit~\cite{Ramette2024}. This gives rise to a directional error pattern concentrated near the seam—the interface between quantum processing units (QPUs) where nonlocal interactions occur.

To study this phenomenon, we adopt a simplified noise model that focuses solely on $X$ errors resulting from faulty Bell states. These bit-flip errors primarily affect the target-side qubits of the teleported CNOTs, leading to stabilizer violations that are spatially localized near one side of the QPU boundary. To capture this effect, the decoder has to account for elevated error rates in this region. Importantly, since $X$ and $Z$ errors propagate independently during non-local gate operations, decoders must address each Pauli error separately, based on the specific noise patterns they observe.


This asymmetric propagation of errors can be intuitively understood using the repetition code. Assuming perfect syndrome extraction within each QPU, the seam qubits effectively behave like a one-dimensional repetition code embedded along the boundary of the two-dimensional color code lattice. This arises because, during stabilizer measurements on boundary polygons, some data qubits lie across the seam in an adjacent QPU. Consequently, Pauli errors can propagate to the boundary qubits. In this picture, the seam acts as a high-noise strip, where the noise model can be adjusted to act differently on distinct sets of qubits. A similar idea of error propagation has been explored in Ref.~\cite{Ramette2024}, where surface code patches are joined along one of their boundaries to study the impact of elevated seam noise on the boundary qubits.

This repetition-code perspective provides a valuable abstraction for modeling asymmetric noise distribution. While we focus on \(X\) errors for simplicity, the framework can be extended to more general noise models. For example, in a realistic setting, the propagation of Pauli \(X\) and \(Z\) errors can be tracked individually, leading to a non-uniform distribution of errors across different qubits. However, analyzing such detailed asymmetries is quite challenging. In this work, we restrict our attention to a simplified scenario in which all seam qubits experience a uniformly scaled bit-flip error relative to the bulk qubits, and we examine its impact on decoder performance in estimating the error threshold.

\section{Simulation Results and Threshold Estimation}\label{plots}

We perform Monte Carlo simulations to estimate the logical error rates of the 6.6.6 color code under a bit-flip noise model. The simulations use two decoding methods: a tensor network-based Matrix Product State (MPS) decoder and a concatenated minimum-weight perfect matching (MWPM) decoder. 

Each simulation trial proceeds as follows:

\begin{itemize}
   
    \item We initialize a triangular color code lattice with code distance $d$.
    \item We apply independent bit-flip errors to each data qubit, with a physical error probability \( p \) for qubits in the bulk and \( p_{seam} \) for those located along the boundary.

    \item We extract the syndrome assuming perfect (noiseless) measurements.
    \item We use the decoders to attempt recovery based on the extracted syndrome.
    \item We classify the trial as a logical failure if the recovered state differs from the initial logical state. 
\end{itemize}

To model the noise asymmetry inherent in distributed architectures, we simulate two distinct noise settings:

\begin{itemize}
    \item \textbf{Uniform (Symmetric) noise model:} All qubits, including seam qubits located at QPU boundaries, experience the same bit-flip error rate $p$.
    
    \item \textbf{Asymmetric noise model:} Seam qubits are subjected to an elevated error rate \( p_{\text{seam}} = \lambda p \), where \( p \) is the physical error rate for bulk qubits and \( \lambda \) is a scaling factor.

\end{itemize}

For each combination of noise model, decoder, code distance, and physical error rate, we perform multiple independent trials. We compute the logical failure probability $p_{\text{fail}}(p, d)$ as the fraction of trials that result in a logical error.

\subsection{Threshold Estimation Using Tensor-Network Based Decoder}

\begin{figure}[hbt!]
    \centering
    \includegraphics[width=\columnwidth]{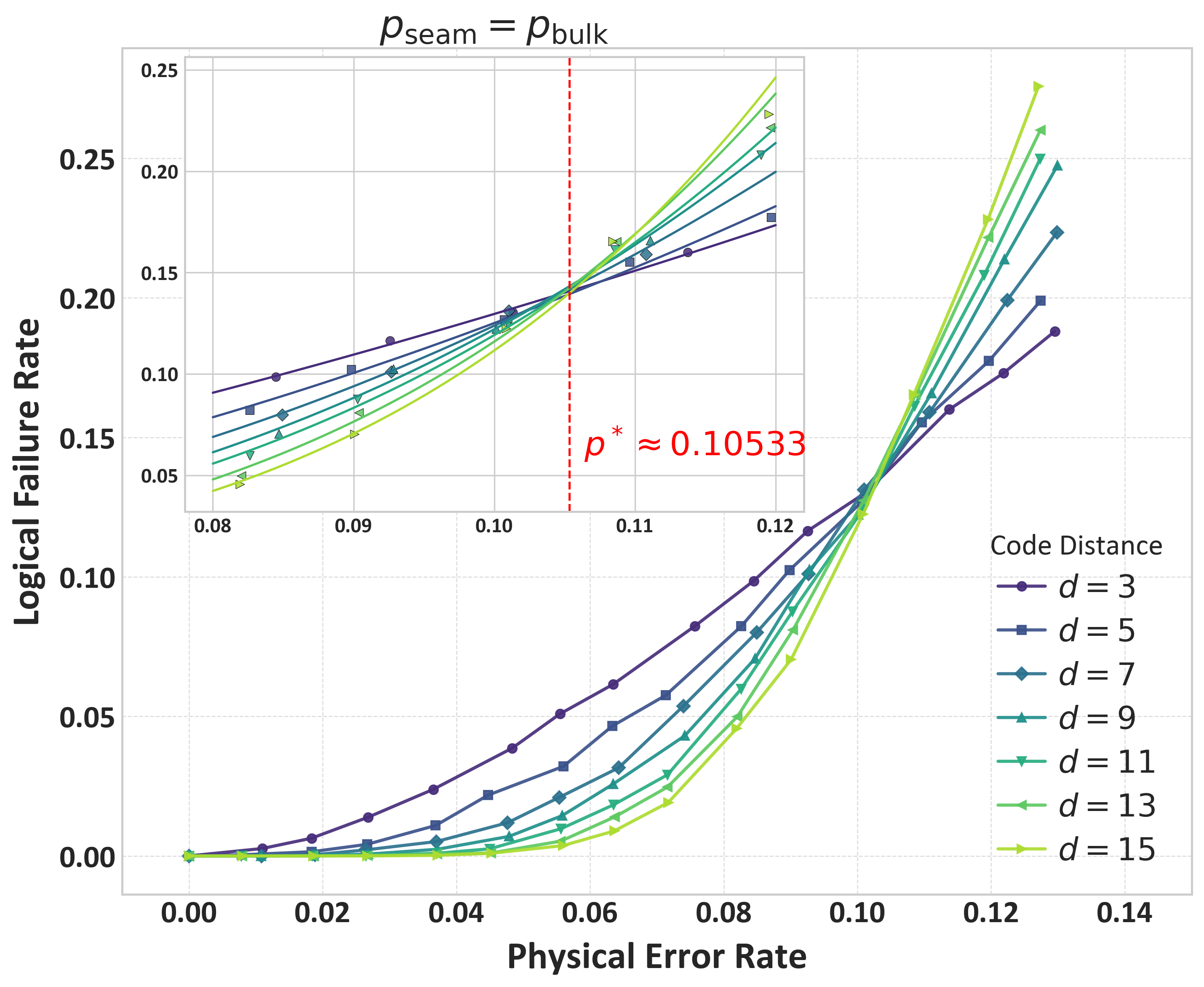}
    \caption{Threshold plot for the 6.6.6 color code under a uniform bit-flip noise model using tensor network based decoder with bond-dimension $\chi=8$. All qubits experience the same physical error rate (p) i.e. $\lambda = 1$. The logical failure probability  is plotted for code distances d = 3, 5, 7, 9,..,15 averaged over 20,000 Monte Carlo trials for each value of $p$.}

    \label{fig:plot5}

\end{figure}

\begin{figure}[hbt!]
    \centering
    \includegraphics[width=\columnwidth]{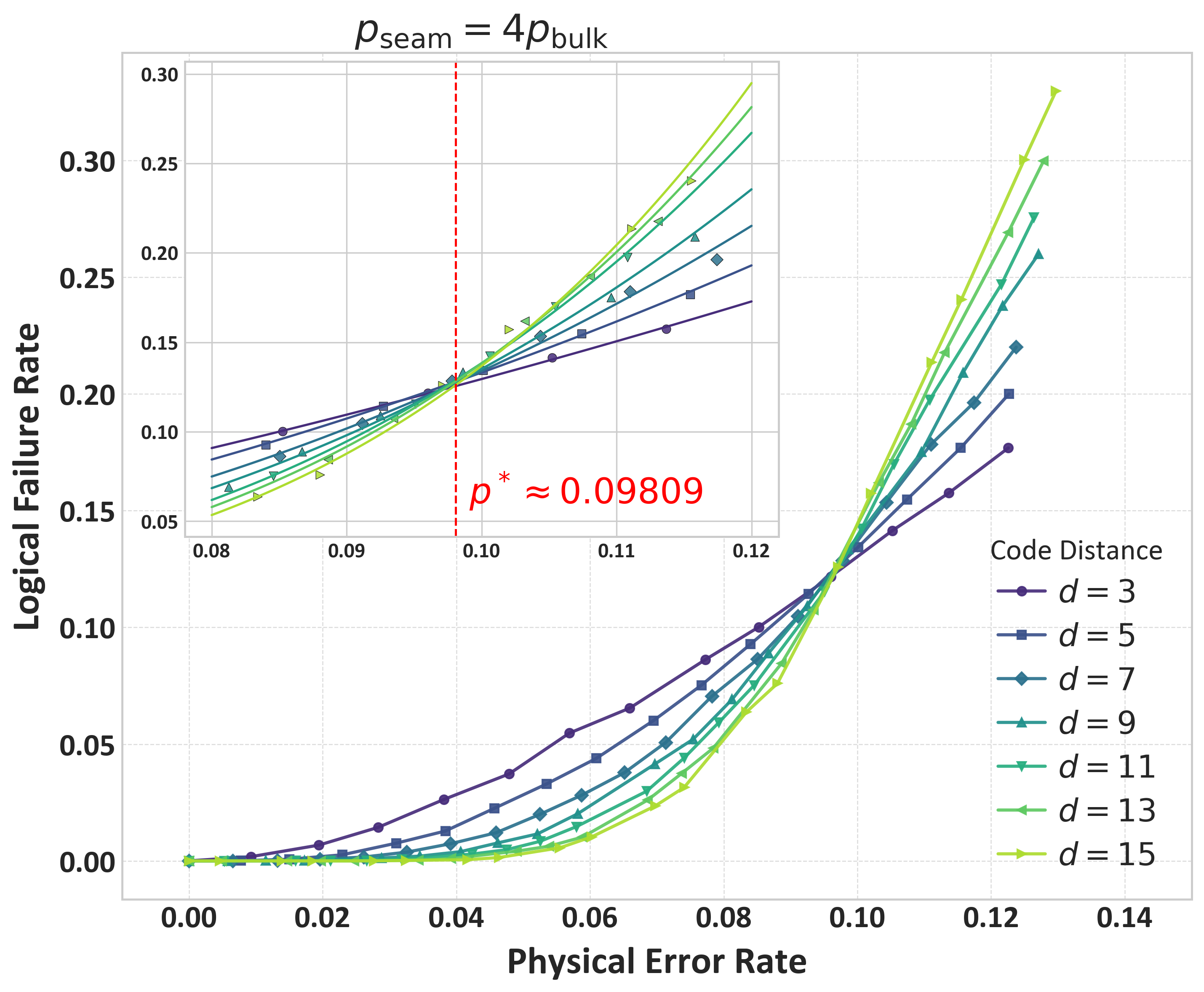}
    \caption{Threshold plot for the 6.6.6 color code under an asymmetric bit-flip noise model using MPS decoder. Seam qubits experience elevated noise ($p_{seam} = 4p$), while bulk qubits have error rate p. Logical failure probabilities are shown for distances d = 3, 5, 7, 9,...,15 using 20,000 Monte Carlo trials for each value of $p$.}

    \label{fig:plot6}

\end{figure}

We analyze the threshold behavior of the 6.6.6 color code under a bit-flip noise model using a tensor network-based decoder based on matrix product states (MPS)~\cite{qecsim} with bond-dimension $\chi=8$. This decoder approximates maximum-likelihood decoding by contracting a tensor network that represents the probability distribution over error configurations conditioned on the syndrome. 



To estimate the threshold \( p^\star \), we fit the logical failure rate \( p_{\text{fail}} \) for each code distance \( d \in \{3,5,7,9,11,13,15\} \) using a linear regression in log-log space over the range \( p \in [0.08, 0.12] \), using the model:
\[
\log_{10}(p_{\text{fail}}) = a_d \log_{10}(p) + b_d.
\]
The intersection point of these linear fits across different distances approximates the threshold---the value of \( p \) at which logical failure becomes approximately distance-independent. Although the logical failure rate follows a power-law scaling with respect to the physical error rate, it exhibits approximately linear behavior near the threshold. Therefore, performing linear regression in log-log space provides an effective way for estimating the threshold value \( p^\star \). 

In the symmetric case ($\lambda = 1$), the decoder yields a threshold estimate of \( p^\star = 0.10533 \pm 0.00190 \) (See Fig.~\ref{fig:plot5}). However, in the asymmetric case with enhanced seam noise, we consider \( \lambda = 4 \) as an illustrative example, under which the threshold drops to
 \( p^\star = 0.09809 \pm 0.00194 \) (See Fig.~\ref{fig:plot6}). The extracted value of \( p^\star \) represents the estimated error threshold. For \( p < p^\star \), increasing the code distance leads to exponential suppression of logical errors, while for \( p > p^\star \), logical errors increase with distance.

The observed reduction in the threshold of the MPS-based decoder under asymmetric noise may be attributed to limitations in its ability to accurately capture the structure of the error distribution in the presence of spatially inhomogeneous noise. Under uniform noise, correlations between qubits tend to remain local and regular, allowing an MPS with moderate bond dimension to approximate the relevant marginal probabilities for effective decoding. However, when seam qubits experience significantly higher error rates than bulk qubits, the resulting spatial bias can introduce more complex correlations—particularly near boundaries and inter-module interfaces. Accurately capturing such a structure typically requires a higher bond dimension ($\chi$), which increases the expressive power of the decoder but also leads to substantially greater computational cost due to the scaling behavior of tensor network contractions~\cite{chubb2021generaltensornetworkdecoding}. In practice, the bond dimension must often remain small to keep decoding feasible, which may result in a less faithful approximation of the true error distribution. This limitation can lead to suboptimal decoding outcomes and a reduction in the effective threshold. These findings suggest that MPS-based decoders may be sensitive to localized, high-noise regions, such as those encountered in modular quantum architectures. While they can provide strong threshold estimates for small code distances, their applicability to large-scale simulations is limited by computational overhead. This motivates the consideration of alternative decoding strategies—such as those based on minimum-weight perfect matching (MWPM) which offer improved scalability, albeit often at the cost of reduced threshold performance.



\subsection{Threshold Estimation using Concatenated MWPM Decoder}

We perform Monte Carlo simulations to evaluate the error-correcting performance of the 6.6.6 color code under a bit-flip noise model using the concatenated minimum-weight perfect matching (MWPM) decoder~\cite{lee2025color}. For each configuration, we simulate \(10^5\) rounds to obtain statistically reliable estimates of the logical failure probability.

The simulations are carried out across a range of physical error rates from \(p = 0.05\) to \(p = 0.09\), in order to capture logical failure behavior both above and below the expected threshold. Code distances from \(d = 3\) to \(d = 45\) are simulated. For threshold fitting, however, we restrict to code distances \(d \geq 7\) and physical error rates within the narrower interval \(p \in [0.078, 0.084]\), where the failure curves become approximately linear near the threshold. This regime allows for accurate and reliable threshold estimation.


We apply the same independent bit-flip noise model in two configurations: first, where the error rate \(p\) is uniformly applied to all qubits, including those at the boundaries of quantum processing units (seam qubits); and second, where seam qubits are subjected to a higher error rate \(p_{\text{seam}} = 4p_{bulk}\), while bulk qubits retain the original error rate \(p\). 

\begin{figure}[hbt!]
    \centering
    \includegraphics[width=\columnwidth]{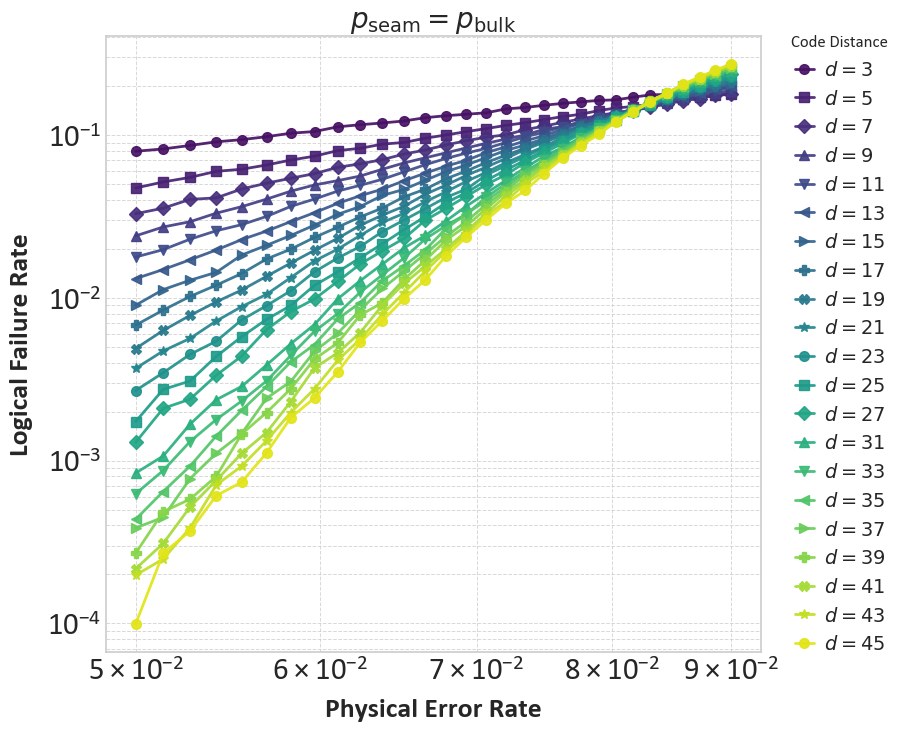}
    \caption{Logical failure rate vs physical error rate under symmetric noise conditions ($p_{\text{seam}} = p$) and the color code distance is varied between 3 and 45.}

    \label{fig:plot7}

\end{figure}

\begin{figure}[hbt!]
    \centering
    \includegraphics[width=\columnwidth]{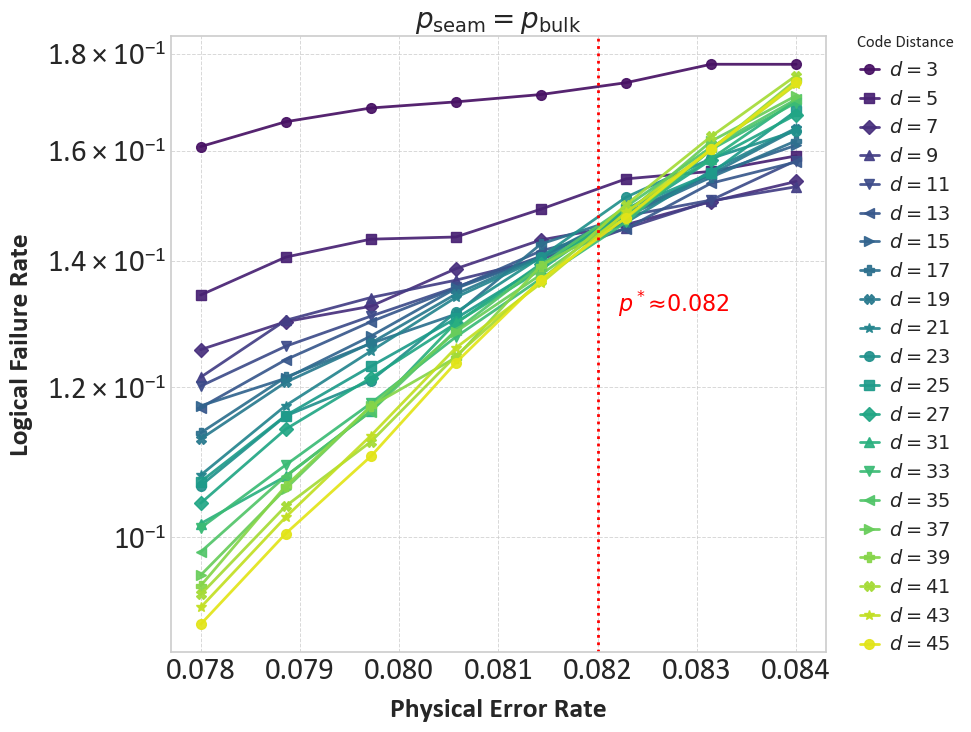}
    \caption{Threshold fitting region for the symmetric noise case ($p_{\text{seam}} = p$), highlighting the linear behavior and convergence of curves  used to extract the threshold.}

    \label{fig:plot8}

\end{figure}


To estimate the threshold, we fit the logical failure rate curves in log-log space using linear regression. For each code distance \(d\), we fit the data using the relation \(\log_{10}(p_{\text{fail}}) = a_d \log_{10}(p) + b_d\). 

\begin{figure}[hbt!]
    \centering
    \includegraphics[width=\columnwidth]{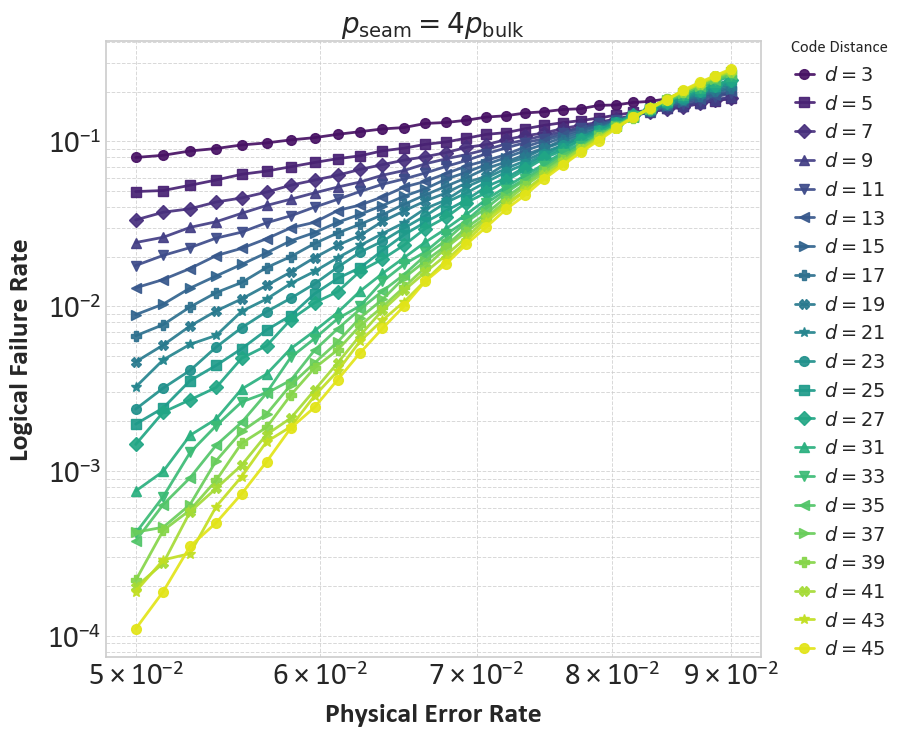}
    \caption{Logical failure rate vs physical error rate under asymmetric noise conditions ($p_{\text{seam}} = 4p$) and the color code distance is varied between 3 and 45.}

    \label{fig:plot9}

\end{figure}

\begin{figure}[hbt!]
    \centering
    \includegraphics[width=\columnwidth]{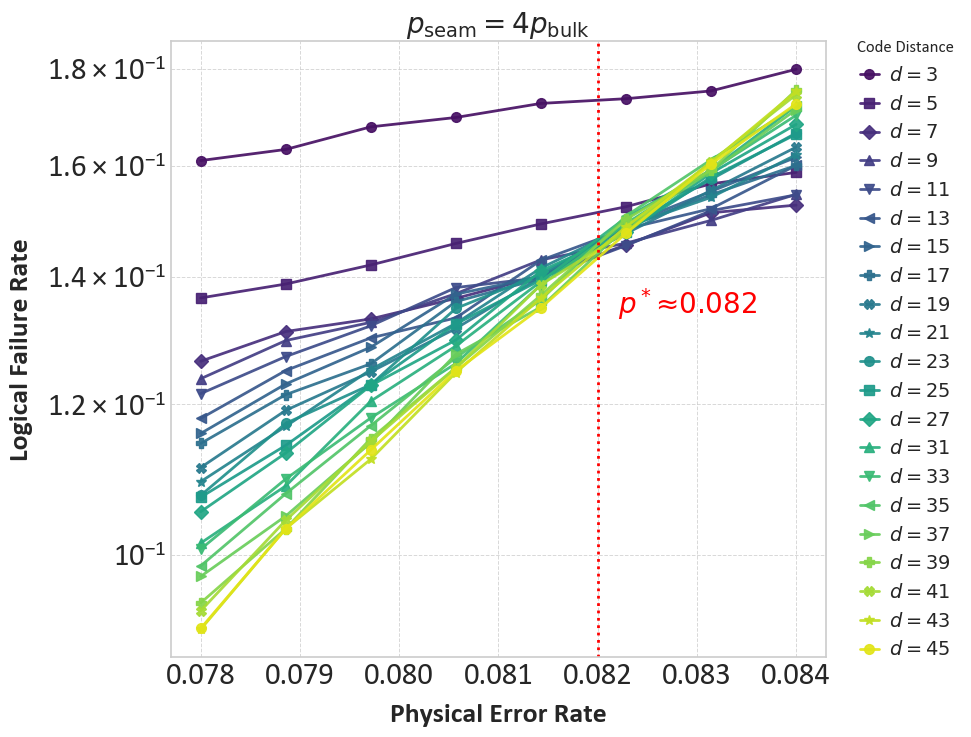}
    \caption{Threshold fitting region for the asymmetric noise case ($p_{\text{seam}} = 4p$), highlighting the linear behavior and convergence of curves used to extract the threshold.}

    \label{fig:plot10}

\end{figure}

In the case where seam and bulk qubits experience symmetric noise, we estimate a threshold of \(p^\star = 0.08205 \pm 0.00114\) (See Fig.~\ref{fig:plot8}). When the seam qubits are assigned a higher error rate \(p_{\text{seam}} = 4p_{bulk}\), the estimated threshold remains close, at \(p^\star = 0.08202 \pm 0.00254\) (See Fig.~\ref{fig:plot10}). Although the central values are nearly identical, we observe slightly increased uncertainty in the asymmetric case. These results suggest that this decoding strategy is well-suited for modular, distributed quantum architectures where seam qubits are naturally more error-prone.

We observe a threshold of approximately 10.53\% under the symmetric noise model and 9.89\% under the fourfold elevated seam noise when using the tensor network decoder. In comparison, the concatenated MWPM decoder achieves a consistent threshold of around 8.2\% across both noise models. Although its threshold is lower than that of the tensor network decoder, it remains stable under asymmetric noise, unlike the tensor network decoder, which shows a slight reduction in performance. This suggests that the concatenated MWPM decoder is more robust to elevated seam noise, offering greater noise tolerance at the cost of a lower threshold. Furthermore, it is computationally more efficient than the tensor network decoder, making it well-suited for large-scale simulations.



\section{Discussion and Conclusion}\label{conclusion}
In this work, we investigated the error correction threshold of the (6.6.6) triangular color code in a distributed quantum computing architecture, where modular patches are interconnected via noisy links. In this setting, seam qubits—located at the boundaries between quantum processing units (QPUs)—are modeled to experience higher error rates than bulk qubits, capturing the asymmetric noise characteristics introduced by imperfect inter-module communication. To analyze the code's performance under such spatially inhomogeneous noise, we evaluated two decoding strategies: a tensor network-based decoder and a concatenated minimum-weight perfect matching (MWPM) decoder. While the tensor network decoder exhibits a modest reduction in threshold, the MWPM-based decoder shows strong robustness, providing effective error correction despite elevated seam noise. These findings highlight the practical potential of color codes for fault-tolerant quantum computation in modular architectures and show their suitability for scalable implementations involving noisy interconnects.

This work serves as an initial step toward understanding the impact of asymmetric noise at modular boundaries on the performance of topological color codes in distributed quantum computing architectures. 
The results and insights presented here can be used to determine the minimum entanglement fidelity required for inter-modular links, as well as the associated networking resource requirements, to meet the error correction threshold in such distributed quantum architectures. Several other compelling directions for future research remain open. One key avenue is the exploration of alternative code families and lattice geometries that may exhibit greater robustness to spatially inhomogeneous noise. There is a need to develop optimal decoders that are not only computationally efficient but also specifically tailored to handle asymmetric noise profiles while maintaining high error thresholds, as encountered in modular and networked quantum systems. Another important direction is the incorporation of more realistic, circuit-level noise models that account for gate errors, state initialization faults, and imperfect measurements. Such models would enable a more accurate and holistic evaluation of decoder performance under experimentally relevant conditions. Furthermore, deriving analytical bounds on the maximum tolerable seam noise for distributed implementations of color codes could provide valuable theoretical insight into the fundamental limits of fault tolerance in modular quantum architectures.

\section{Acknowledgments}
 NKC and KPS gratefully acknowledge support from Cisco Systems, Inc.. This research was supported in part by the University of Pittsburgh Center for Research Computing and Data, \texttt{RRID:SCR\_022735}, through the resources provided. Specifically, this work used the HTC cluster, which is supported by NIH award number \texttt{S10OD028483}.

\bibliographystyle{IEEEtran}
\bibliography{references}

\end{document}